\documentclass[12pt]{article}
\usepackage{mathptmx}
\usepackage[utf8]{inputenc}
\usepackage[letterpaper, portrait, margin=1in]{geometry}
\usepackage[english]{babel}
\usepackage{amsmath, amssymb}
\usepackage{graphicx}
\usepackage{enumitem}
\usepackage{ragged2e}
\usepackage[range-units=brackets, range-phrase=-]{siunitx}
\usepackage[symbol]{footmisc}
\usepackage{multicol}
\usepackage{cite}
\usepackage{xcolor}
\definecolor{dullpurple}{rgb}{0.431,0.188,0.534}
\definecolor{darkgreen}{rgb}{0.075,0.302,0.047}
\definecolor{dullred}{rgb}{0.706,0.208,0.192}
\usepackage[colorlinks=true, urlcolor=dullpurple, citecolor=dullred, linkcolor=darkgreen]{hyperref}
\usepackage[babel=true]{microtype}

\DeclareSIUnit{\parsec}{pc}
\DeclareSIUnit{\Mpc}{\mega\parsec}
\DeclareSIUnit{\h}{\mathit{h}}

\makeatletter
\renewcommand\section{\@startsection{section}{1}{\z@}%
	{-2.6ex \@plus -0.7ex \@minus -.4ex}%
	{1.5ex \@plus.4ex \@minus .4ex}%
	{\normalfont\large\bfseries}}
\makeatother

\newcommand{\ANLHEP}{HEP Division, Argonne National Laboratory, Lemont, IL~60439, USA}
\newcommand{\APC}{Laboratoire Astroparticule et Cosmologie (APC), CNRS/IN2P3, Universit\'e Paris Diderot, 75205~Paris, France}

\newcommand{\BenGurion}{Department of Physics, Ben-Gurion University, Be'er~Sheva~84105, Israel}
\newcommand{\BNL}{Brookhaven National Laboratory, Upton, NY~11973, USA}
\newcommand{\Brown}{Brown University, Providence, RI~02912, USA}

\newcommand{\BU}{Boston University, Boston, MA~02215, USA}

\newcommand{\Caltech}{California Institute of Technology, Pasadena, CA~91125, USA}

\newcommand{\Cavendish}{Astrophysics Group, Cavendish Laboratory, University of Cambridge, Cambridge~CB3~0HE, UK}
\newcommand{\CCA}{Center for Computational Astrophysics, Flatiron Institute, New York, NY~10010, USA}

\newcommand{\CfA}{Harvard-Smithsonian Center for Astrophysics, Cambridge, MA~02138, USA}

\newcommand{\Cincinnati}{University of Cincinnati, Cincinnati, OH~45221, USA}
\newcommand{\CITA}{Canadian Institute for Theoretical Astrophysics, University of Toronto, Toronto, ON~M5S~3H8, Canada}

\newcommand{\Columbia}{Columbia University, New York, NY~10027, USA}
\newcommand{\Cornell}{Cornell University, Ithaca, NY~14853, USA}

\newcommand{\CWRU}{Case Western Reserve University, Cleveland, OH~44106, USA}
\newcommand{\daa}{Department of Astronomy and Astrophysics, University of Toronto, ON~M5S~3H4, Canada}
\newcommand{\damtp}{DAMTP, University of Cambridge, Cambridge CB3~0WA, UK}

\newcommand{\DFI}{Departamento de F\'isica, FCFM, Universidad de Chile, Santiago, Chile}

\newcommand{\dunlap}{Dunlap Institute for Astronomy and Astrophysics, University of Toronto, ON~M5S~3H4, Canada}
\newcommand{\Durham}{Department of Physics, Durham University, Durham DH1~3LE, UK}
\newcommand{\ED}{University of Edinburgh, Edinburgh EH8~9YL, UK}
\newcommand{\EPFL}{Institute of Physics, Laboratory of Astrophysics, École Polytechnique Fédérale de Lausanne (EPFL), Observatoire de Sauverny, 1290~Versoix, Switzerland}

\newcommand{\GSFC}{Goddard Space Flight Center, Greenbelt, MD~20771, USA}

\newcommand{\HarvardPhys}{Department of Physics, Harvard University, Cambridge, MA~02138, USA}

\newcommand{\HKUST}{The Hong Kong University of Science and Technology, Hong~Kong~SAR, China}

\newcommand{\IAP}{Institut d'Astrophysique de Paris (IAP), CNRS \& Sorbonne University, 75014~Paris, France}
\newcommand{\IAS}{Institute for Advanced Study, Princeton, NJ~08540, USA}

\newcommand{\IFPU}{Institute for Fundamental Physics of the Universe (IFPU), 34014~Trieste, Italy}

\newcommand{\IFUNAM}{Instituto de F\'{i}sica (IFUNAM), Universidad Nacional Aut\'onoma de M\'exico, 04510~Ciudad de México, Mexico}

\newcommand{\Imperial}{Theoretical Physics, Blackett Laboratory, Imperial College, London SW7~2AZ, UK}

\newcommand{\INFN}{National Institute for Nuclear Physics (INFN), 34127~Trieste, Italy}
\newcommand{\INFNFE}{Istituto Nazionale di Fisica Nucleare, Sezione di Ferrara, 40122~Ferrara, Italy}

\newcommand{\INFNRM}{Istituto Nazionale di Fisica Nucleare, Sezione di Roma, 00185~Roma, Italy}

\newcommand{\ioa}{Institute of Astronomy, University of Cambridge, Cambridge CB3~0HA, UK}

\newcommand{\IPMU}{Kavli Institute for the Physics and Mathematics of the Universe (WPI), University of Tokyo, 277-8583~Kashiwa, Japan}

\newcommand{\ITFA}{Institute for Theoretical Physics, University of Amsterdam, 1098~XH~Amsterdam, The Netherlands}

\newcommand{\JHU}{Johns Hopkins University, Baltimore, MD~21218, USA}

\newcommand{\JPL}{Jet Propulsion Laboratory, California Institute of Technology, Pasadena, CA~91109, USA}
\newcommand{\KASSI}{Korea Astronomy and Space Science Institute, Daejeon~34055, Korea}
\newcommand{\kavli}{Kavli Institute for Cosmology, Cambridge CB3~0HA, UK}

\newcommand{\KICP}{Kavli Institute for Cosmological Physics, Chicago, IL~60637, USA}
\newcommand{\KIPAC}{Kavli Institute for Particle Astrophysics and Cosmology, Stanford, CA~94305, USA}

\newcommand{\KSU}{Kansas State University, Manhattan, KS~66506, USA}

\newcommand{\LBL}{Lawrence Berkeley National Laboratory, Berkeley, CA~94720, USA}

\newcommand{\LLNL}{Lawrence Livermore National Laboratory, Livermore, CA~94550, USA}

\newcommand{\Melbourne}{School of Physics, The University of Melbourne, Parkville, VIC~3010, Australia}

\newcommand{\MIT}{Massachusetts Institute of Technology, Cambridge, MA~02139, USA}

\newcommand{\NAOC}{National Astronomical Observatories, Chinese Academy of Sciences, Beijing, China}
\newcommand{\NCBJ}{National Center for Nuclear Research, 02-093~Warsaw, Poland}

\newcommand{\OSU}{The Ohio State University, Columbus, OH~43212, USA}
\newcommand{\OU}{Department of Physics and Astronomy, Ohio University, Athens, OH~45701, USA}
\newcommand{\OskarKlein}{Oskar Klein Centre for Cosmoparticle Physics, Stockholm University, AlbaNova, 106~91~Stockholm, Sweden}
\newcommand{\Oxford}{University of Oxford, Oxford OX1~3RH, UK}

\newcommand{\ParisSud}{LAL, Universit\'{e} Paris-Sud, 91898~Orsay~Cedex, France \& CNRS/IN2P3, 91405~Orsay, France}
\newcommand{\PI}{Perimeter Institute, Waterloo, ON~N2L~2Y5, Canada}

\newcommand{\Port}{Institute of Cosmology \& Gravitation, University of Portsmouth, Portsmouth PO1~3FX, UK}
\newcommand{\Princeton}{Princeton University, Princeton, NJ~08544, USA}

\newcommand{\QMUL}{Queen Mary University of London, London E1~4NS, UK}

\newcommand{\Rice}{Department of Physics \& Astronomy, Rice University, Houston, TX~77005, USA}

\newcommand{\RomaS}{Dipartimento di Fisica, Universit\`{a} La Sapienza, 00185~Roma, Italy}

\newcommand{\Sejong}{Department of Physics and Astronomy, Sejong University, Seoul~143-747, Korea}

\newcommand{\SHAO}{Shanghai Astronomical Observatory (SHAO), Shanghai~200030, China}

\newcommand{\SISSA}{International School for Advanced Studies~(SISSA), 34136~Trieste, Italy}
\newcommand{\SimonFraser}{Department of Physics, Simon Fraser University, Burnaby, BC~V5A~1S6, Canada}
\newcommand{\SLAC}{SLAC National Accelerator Laboratory, Menlo Park, CA~94025, USA}
\newcommand{\SMU}{Southern Methodist University, Dallas, TX~75275, USA}

\newcommand{\SoCal}{University of Southern California, Los Angeles, CA~90089, USA}
\newcommand{\Stanford}{Stanford University, Stanford, CA~94305, USA}
\newcommand{\StonyBrook}{Stony Brook University, Stony Brook, NY~11794, USA}

\newcommand{\SussexAstronomy}{Astronomy Centre, School of Mathematical and Physical Sciences, University of Sussex, Brighton BN1~9QH, UK}
\newcommand{\Syracuse}{Syracuse University, Syracuse, NY~13244, USA}

\newcommand{\TIFR}{Tata Institute of Fundamental Research, Mumbai~400005, India}

\newcommand{\UAM}{Universidad Aut\'onoma de Madrid, 28049~Madrid, Spain}

\newcommand{\UCBP}{Department of Physics, University of California Berkeley, Berkeley, CA~94720, USA}
\newcommand{\UCBSSL}{Space Sciences Laboratory, University of California Berkeley, Berkeley, CA~94720, USA}
\newcommand{\UCD}{University of California Davis, Davis, CA~95616, USA}
\newcommand{\UChicago}{University of Chicago, Chicago, IL~60637, USA}
\newcommand{\UCI}{University of California Irvine, Irvine, CA~92697, USA}
\newcommand{\UCLA}{University of California Los Angeles, Los Angeles, CA~90095, USA}
\newcommand{\UCL}{University College London, London WC1E~6BT, UK}
\newcommand{\UCR}{University of California Riverside, Riverside, CA~92521, USA}

\newcommand{\UCSD}{University of California San Diego, La Jolla, CA~92093, USA}
\newcommand{\UFL}{University of Florida, Gainesville, FL~32611, USA}

\newcommand{\UGTO}{Divisi\'on de Ciencias e Ingenier\'ias, Universidad de Guanajuato, Le\'on~37150, M\'exico}

\newcommand{\UMich}{University of Michigan, Ann Arbor, MI~48109, USA}
\newcommand{\UMN}{University of Minnesota, Minneapolis, MN 55455, USA}
\newcommand{\UnB}{Instituto de F\'{i}sica, Universidade de Bras\'{i}lia, 70919-970~Bras\'{i}lia, Brazil}

\newcommand{\UNIPD}{Dipartimento di Fisica e Astronomia ``G. Galilei'', Universit\`a degli Studi di Padova, 35131~Padova, Italy}
\newcommand{\UNM}{University of New Mexico, Albuquerque, NM~87131, USA}

\newcommand{\UoM}{Jodrell Bank Center for Astrophysics, School of Physics and Astronomy, University of Manchester, Manchester M13~9PL, UK}
\newcommand{\UPenn}{Department of Physics and Astronomy, University of Pennsylvania, Philadelphia, PA~19104, USA}
\newcommand{\UR}{Department of Physics and Astronomy, University of Rochester, Rochester, NY~14627, USA}
\newcommand{\UrbanaC}{Department of Physics, University of Illinois at Urbana-Champaign, Urbana, IL~61801, USA}

\newcommand{\UTD}{University of Texas at Dallas, Richardson, TX~75080, USA}

\newcommand{\UWaterloo}{Department of Physics and Astronomy, University of Waterloo, Waterloo, ON~N2L~3G1, Canada}
\newcommand{\UWMadison}{Department of Physics, University of Wisconsin-Madison, Madison, WI~53706, USA}

\newcommand{\VSI}{Van Swinderen Institute for Particle Physics and Gravity, University of Groningen, 9747~AG~Groningen, The~Netherlands}
\newcommand{\VT}{Virginia Tech, Blacksburg, VA~24061, USA}

\newcommand{\WCA}{Centre for Astrophysics, University of Waterloo, Waterloo, ON~N2L~3G1, Canada}

\newcommand{\WVU}{CSEE, West Virginia University, Morgantown, WV~26505, USA}
\newcommand{\WVUGWAC}{Center for Gravitational Waves and Cosmology, West Virginia University, Morgantown, WV~26505, USA}
\newcommand{\Wyoming}{Department of Physics and Astronomy, University of Wyoming, Laramie, WY~82071, USA}
\newcommand{\Yale}{Department of Physics, Yale University, New Haven, CT~06520, USA}

\begin{document}
\pagenumbering{roman}
\thispagestyle{empty}

{\raggedright 
\huge
Astro2020 Science White Paper \linebreak

Scratches from the Past:\\Inflationary Archaeology through Features in\\[6pt]the Power Spectrum of Primordial Fluctuations \linebreak
\normalsize

\noindent \textbf{Thematic Areas:} Cosmology and Fundamental Physics \linebreak

\textbf{Principal Author:}

Name: An\v{z}e Slosar	
\linebreak
Institution: Brookhaven National Laboratory
\linebreak
Email: \texttt{anze@bnl.gov}
\linebreak
Phone: (631) 344 8012
\linebreak

\textbf{Co-authors:} See next page.
\linebreak

\noindent\textbf{Abstract:}
Inflation may provide unique insight into the physics at the highest available energy scales that cannot be replicated in any realistic terrestrial experiment. Features in the primordial power spectrum are generically predicted in a wide class of models of inflation and its alternatives, and are observationally one of the most overlooked channels for finding evidence for non-minimal inflationary models. Constraints from observations of the cosmic microwave background cover the widest range of feature frequencies, but the most sensitive constraints will come from future large-scale structure surveys that can measure the largest number of linear and quasi-linear modes.
}

\newpage
{\raggedright 
\textbf{Authors\hskip1pt/\hskip1ptEndorsers\footnote[1]{Names in bold indicate significant contribution.}:}
{
Kevork N.\ Abazajian$^{1}$, 
Muntazir Abidi$^{2}$, 
Peter Adshead$^{3}$, 
Zeeshan Ahmed$^{4}$, 
David Alonso$^{5}$, 
Mustafa A.\ Amin$^{6}$, 
Behzad Ansarinejad$^{7}$, 
Robert Armstrong$^{8}$, 
Carlo Baccigalupi$^{9,10,11}$, 
Kevin Bandura$^{12,13}$
% ...
}
\linebreak
\begin{raggedright}
\setlength{\columnsep}{18pt}
\begin{multicols}{2}
\footnotesize

\noindent  $^{1}$ \UCI \\
$^{2}$ \damtp \\
$^{3}$ \UrbanaC \\
$^{4}$ \SLAC \\
$^{5}$ \Oxford \\
$^{6}$ \Rice \\
$^{7}$ \Durham \\
$^{8}$ \LLNL \\
$^{9}$ \SISSA \\
$^{10}$ \IFPU \\
$^{11}$ \INFN \\
$^{12}$ \WVU \\
$^{13}$ \WVUGWAC \\
% ...

\normalsize
\end{multicols}
\end{raggedright}
}

\clearpage
\pagenumbering{arabic}

\section{Introduction}

The standard cosmological model has proven to be incredibly successful and has been confirmed repeatedly over several generations of improving cosmic microwave background~(CMB) and large-scale structure~(LSS) experiments. A crucial part of this model are the initial seed fluctuations which are Gaussian with a nearly scale-invariant power spectrum. While these are generic predictions of inflation, current data does not point to a specific mechanism. More complex models of inflation can imprint features in the primordial spectra (see e.g.~\cite{Chen:2010xka, Chluba:2015bqa} for reviews) which, if found, would be a groundbreaking discovery that would open an entirely new window into the primordial universe. This science white paper argues that such features are \emph{generic} in many classes of models of inflation and its alternatives, and are worth a dedicated effort to find them. We therefore argue for support of a new generation of experiments surveying both the~CMB and the~LSS, with the goal of maximizing the range in spatial scales and the total number of accessible linear and quasi-linear modes.

\section{Motivation and Theoretical Overview}

All structure in the universe originated from the dynamics of fields in the very early universe, prior to the moment when the Standard Model particles thermalized. During this era, density fluctuations were spontaneously created from the fluctuations of one or many degrees of freedom that were relevant at that time. Single-field slow-roll inflation is one such possibility that is currently consistent with observations. In this case, the exponential expansion of the universe is responsible for converting vacuum fluctuations of the inflaton into macroscopic classical density perturbations.

The space of inflationary models (and their alternatives) is vast and includes a number of scenarios where the dynamics that give rise to the primordial density fluctuations are more complicated than in single-field inflation. The early universe would have involved many degrees of freedom with complicated interactions, leading to a variety of non-adiabatic or even classical production mechanisms. These dynamics can also give rise to an excited state for the degrees of freedom and significantly alter the description of this era in cosmic history. Any of these effects may leave a residual sharp feature in the initial conditions of the hot big bang. 

Broadly speaking, features in the primordial spectra are rooted in one of the most fundamental challenges in inflationary model building: creating a flat potential or, more generally, making the slow-roll parameters small. While one can arrive at such a model by introducing a new symmetry, these very symmetries are known to be broken in a theory of quantum gravity. While such effects are known to be irrelevant for earthly phenomena, inflation is famously sensitive to them (see e.g.~\cite{Baumann:2009ds, Baumann:2014nda}). Models which avoid the most drastic effects of quantum gravity can still have relics of this basic tension in various sub-leading violations of scale invariance in the form of features. Detecting such features would provide a unique insight into the physics of the primordial universe. In addition, it could provide evidence for particular models of inflation or one of its alternatives, or identify the existence of new particles and forces in the early universe.

For the purpose of observations, primordial features are characterized by density perturbations that contain some small components that significantly depart from scale invariance. These signatures arise in broad classes of models, including both inflation and its alternatives. There are several general types of feature models which we classify according to their underlying generation mechanisms and illustrate in the left panel of Fig.~\ref{fig:features}:
\begin{figure}[t]
\centering
\includegraphics[trim=0 0 0 6]{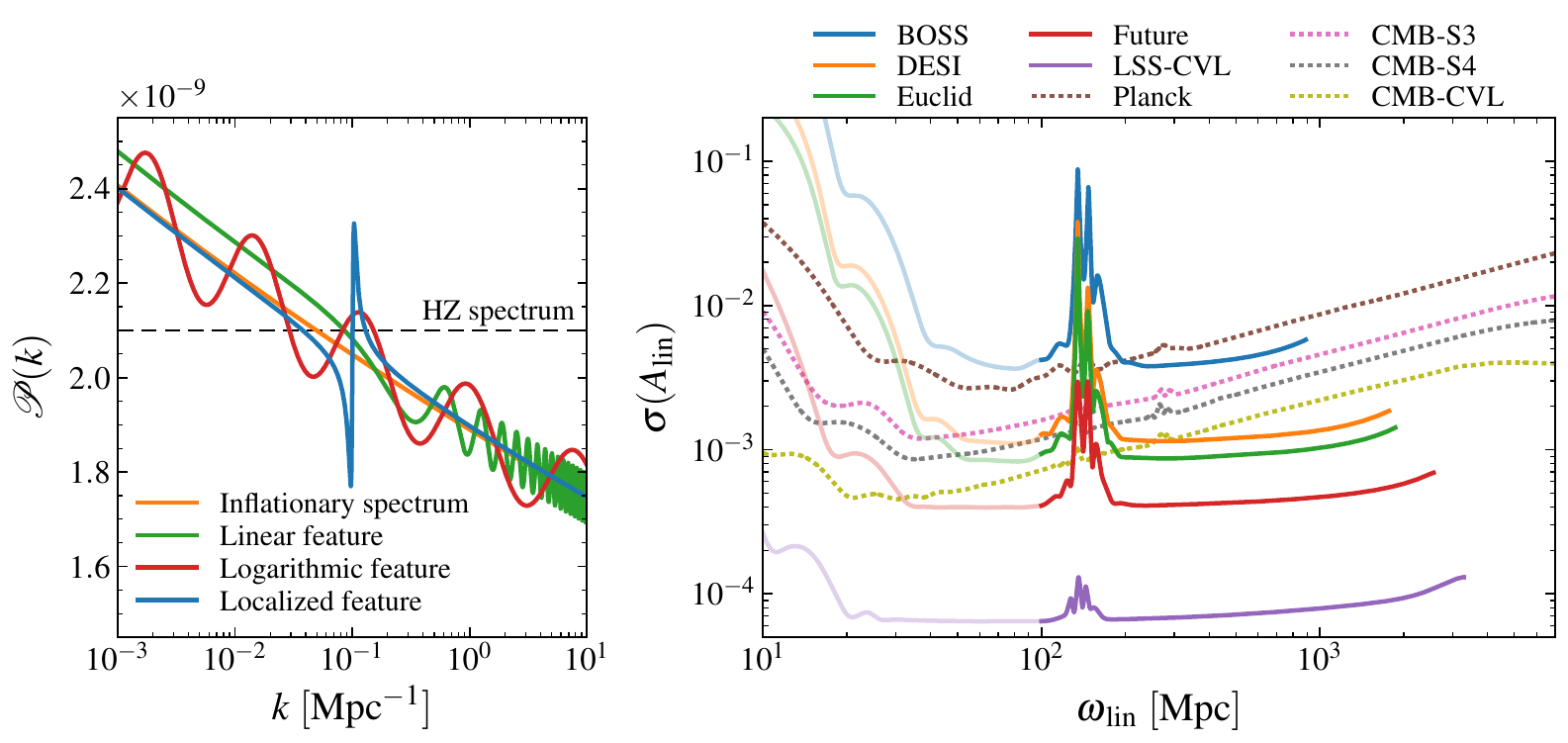}\vspace{-10pt}
\caption{\textit{Left:}~Schematic illustration of the dimensionless power spectrum of primordial curvature fluctuations,~$\mathcal{P}\hskip-1pt(k)$. The almost scale-invariant (inflationary) power spectrum consistent with current CMB~data ($n_\mathrm{s} = 0.965$)~\cite{Akrami:2018odb} is displayed together with three different types of models: sharp and resonant features imprint additional oscillations which are linear and logarithmic in the wavenumber, whereas localized features only depart from the power-law behavior around a distinct wavenumber. For comparison, we also show the Harrison-Zel'dovich~(HZ) spectrum, which is perfectly scale invariant and corresponds to an infinitely slowly-rolling scalar field.
\textit{Right:}~Forecasted sensitivity for the ``feature spectrometer'' (adapted from~\cite{Beutler:2019prep}). The potential reach of various CMB~(dotted) and LSS~(solid)~surveys to constrain the amplitude of linear features,~$A_\mathrm{lin}$, is presented as a function of their frequency~$\omega_{\hskip1pt\mathrm{lin}}$ (for the $\phi_{\hskip0.5pt\mathrm{lin}}=\pi/2$ mode). The modeling of the galaxy power spectrum and the experimental specifications are essentially the same as in~\cite{Baumann:2017gkg}, with cosmic variance-limited~(CVL) observations of the~CMB up to $\ell_\mathrm{max}^T = 3000$ and $\ell_\mathrm{max}^P = 5000$, and of LSS up to $z_\mathrm{max}=6$ and $k_\mathrm{max}=\SI{0.75}{\h\per\Mpc}$. The LSS~forecasts with $\omega_{\hskip1pt\mathrm{lin}} \lesssim \SI{100}{\Mpc}$ should be treated cautiously since these low frequencies are more sensitive to the details of signal modeling. We refer to~\cite{Beutler:2019prep} for details. 
}\vspace{-4pt}
\label{fig:features}
\end{figure}

\vspace{-0.1cm}
\begin{itemize}[leftmargin=0.8cm]
\setlength\itemsep{-0.05cm}
\item \emph{Resonant feature (oscillations in~$\log k$)}.
The background evolution in this class of models oscillates around the attractor solution with a frequency that is larger than the horizon scale. This background oscillation resonates with the quantum modes of the density perturbations and generates a scale-dependent oscillatory component in the density perturbations~\cite{Chen:2008wn}. A well-known example is the axion monodromy model in the inflationary scenario~\cite{Silverstein:2008sg, Flauger:2009ab}, in which case the phase of the resonant feature as a function of the wavenumber~$k$ behaves as $\cos \left(\Omega \log(2k) + \phi\right)$, where~$\Omega$ and~$\phi$ are constants.

\item \emph{Sharp feature (oscillations in~$k$)}.
Models in this class temporarily deviate from the attractor solution at some point during their evolution~\cite{Starobinsky:1992ts}. The deviation can have a variety of physical origins and is generally referred to as a sharp feature~\cite{Starobinsky:1992ts, Adams:2001vc, Bean:2008na, Achucarro:2010da, Miranda:2012rm, Bartolo:2013exa, Hazra:2014goa}. This type of feature induces an oscillatory component in the primordial power spectrum whose phase as a function of~$k$ behaves as $\cos(2k/k_f + \phi)$, where~$k_f$ and~$\phi$ are approximately constants. This model-independent sinusoidal running has a highly model-dependent envelop. We note that there is no a-priori reason to assume a single sharp feature since a periodic~\cite{Green:2009ds} or random~\cite{Green:2014xqa, Amin:2015ftc, Garcia:2019icv} distribution of features may also be generic. Having said that, in most cases, one can equivalently treat the features as a sum of oscillations or local structures in~$k$. In some special cases, the first bump or dip of the sharp feature signal is much more significant than the rest of the oscillations~\cite{Barnaby:2009dd, Chantavat:2010vt} and a template with such a distinct signature in the primordial spectra may be more practical.

\item \emph{Primordial Standard Clocks}.
Massive fields in the primordial universe oscillate either classically~\cite{Chen:2011zf, Chen:2014joa} or quantum mechanically~\cite{Chen:2015lza, Chen:2018cgg}. These oscillations work as standard clocks~and imprint clock signals in the density perturbations. The phase of this oscillatory signal as a function of wavenumber directly records the scale factor of the universe as a function of time,~$a(t)$~\cite{Chen:2011zf, Chen:2014joa, Chen:2015lza, Chen:2018cgg}. Since this function is the defining property, a measurement would provide direct evidence for such a scenario of the primordial universe, whether it is inflation or one of its alternatives.
\end{itemize}
\vspace{-0.3cm}
All these types of features in the power spectrum have correlated signals in non-Gaussianities, i.e.~higher-point statistics, which can be used as further supportive evidence~\cite{Chen:2008wn, Flauger:2010ja, Chen:2010bka, Achucarro:2012fd, Gong:2014spa, Chen:2014cwa, Palma:2014hra}.

The physics responsible for these scenarios is often deeply tied to the fundamental origin of the respective model. Let us illustrate this with the first class of examples. Axion fields are appealing inflaton candidates because of their underlying shift symmetry in field space. In string theory, or in the presence of multiple interacting axions, this discrete shift symmetry is generically broken, leading to a field range larger than the period of the underlying axion potential. Under these conditions, the small underlying axion period imprints oscillatory features in the power spectrum and higher-order statistics of the scalar perturbations. The amplitude and precise shape of these features is model-dependent; its period may drift with time during inflation, for instance, which requires careful analyses~\cite{Flauger:2014ana}. If the inflaton couples to other degrees of freedom, those may be periodically produced at a mass scale~$\mu$ up to the scale of the inflaton kinetic energy density, $\mu^2\sim \dot\phi$. This represents a reach of observations to a scale higher than the inflationary Hubble scale~\cite{Flauger:2016idt}. From specific examples like these, which are interesting in their own right, we can extract broader lessons for low-energy effective field theory and data analysis.

In the most general case, features represent any component that modulates a smooth ``back-ground'' given by a near power-law power spectrum produced by slow-roll, $\mathcal{P}_0(k) = A_\mathrm{s}\,(k/k_*)^{n_\mathrm{s}-1}$, with scalar amplitude~$A_\mathrm{s}$, scalar spectral index~$n_\mathrm{s}$ and pivot scale~$k_*$. Some of these models are localized in Fourier space, e.g.~those generated by kinks or other local features in the inflationary potential, others oscillate with a sufficiently high frequency to be distinguishable from the smooth component. As discussed above, two archetype models are linear oscillations,\vspace{-5pt}
\begin{equation}
\mathcal{P}\hskip-1pt(k) = \mathcal{P}_0(k) \left[1+ A_\mathrm{lin}\cos(\omega_{\hskip1pt\mathrm{lin}}\, k + \phi_{\hskip0.5pt\mathrm{lin}}) \right],\vspace{-5pt}
\end{equation}
which modulate the minimal slow-roll power-law spectrum by a sinusoidal fluctuation with a certain relative amplitude~$A_\mathrm{lin}$, frequency~$\omega_{\hskip1pt\mathrm{lin}}$ and phase~$\phi_{\hskip0.5pt\mathrm{lin}}$, and logarithmic oscillations,\vspace{-5pt}
\begin{equation}
\mathcal{P}\hskip-1pt(k) = \mathcal{P}_0(k) \left[1+ A_\mathrm{log}\cos\left(\omega_{\hskip1pt\mathrm{log}} \log(k/k_*) + \phi_{\hskip0.5pt\mathrm{log}}\right) \right],\vspace{-5pt}
\end{equation}
with the same three parameters. However, the details can vary significantly: possible runnings of the frequency~\cite{Flauger:2014ana}, locality of the feature~\cite{Starobinsky:1992ts, Adams:2001vc, Bean:2008na, Achucarro:2010da, Miranda:2012rm, Bartolo:2013exa, Achucarro:2013cva}, and features which mix properties of the sharp and resonant scenarios~\cite{Chen:2014joa, Chen:2014cwa} are possibilities within the vast landscape of models.

Various approaches exist in the literature for reconstructing the primordial power spectrum (e.g.~\cite{Bridle:2003sa, Guo:2011re, Aich:2011qv, Hazra:2016fkm}). For models with well-specified functional forms, including the logarithmic and linear oscillations, these additional features are typically incorporated directly into a typical power spectrum analysis. In the case of axion monodromy inflation, for instance, a slow drifting of~the frequency and phase of the logarithmic oscillations is expected and can be included in the analysis. In the absence of a model, linear oscillations can be a useful basis in which to look for features, as these oscillations form an orthogonal basis of functions on a given range of wavenumbers, much like a time-series analysis problem. Furthermore, numerous non-parametric reconstruction techniques of the primordial power spectrum have also been developed, including penalized likelihood reconstruction~\cite{Gauthier:2012aq, Planck:2013jfk, Ade:2015lrj, Akrami:2018odb}, Bayesian reconstruction~\cite{Vazquez:2012ux, Aslanyan:2014mqa, Finelli:2016cyd, Planck:2013jfk, Ade:2015lrj, Akrami:2018odb}, cubic spline reconstruction~\cite{Planck:2013jfk, Ade:2015lrj, Akrami:2018odb}, Richardson-Lucy reconstruction~\cite{Shafieloo:2003gf, Hazra:2014jwa}, generalised slow-roll methods~\cite{Kadota:2005hv, Dvorkin:2009ne, Hu:2011vr} and principle component analysis~\cite{Leach:2005av, Dvorkin:2010dn, Dvorkin:2011ui}.

\section{Current and Future Observational Trends}
Any features in the primordial power spectrum will result in features in all observables that are sensitive to fluctuations in the universe. Employing different observables is useful since they probe complementary scales and have different advantages.

\vspace{3pt}
\textbf{Cosmic Microwave Background} anisotropies are the cornerstone of most cosmological analyses, including the search for features. The advantages of the CMB are that (i)~it probes the largest accessible scales, (ii)~the physics is entirely linear and, therefore, under complete theoretical control, and (iii)~it is extremely well measured. The main disadvantages are that projection and transfer-function effects, i.e.~the linear transformation between the primordial power spectrum and the observed spherical power spectrum~$C_\ell$, can smear high-frequency oscillations. Moreover, the temperature power spectrum has been measured to the cosmic variance limit up to $\ell\sim1600$~\cite{Aghanim:2015xee} and the future will therefore only bring relatively incremental improvements (factors of a few at most) as the measurements in the polarization signal become cosmic variance-limited (see e.g.~\cite{Finelli:2016cyd, Hazra:2017joc, Sohn:2019rlq, Beutler:2019prep}).

Current searches in the~CMB have not found any significant detection (cf.~e.g.~\cite{Pahud:2008ae, Adshead:2011jq, Meerburg:2011gd, Peiris:2013opa, Meerburg:2013cla, Meerburg:2013dla, Easther:2013kla, Miranda:2013wxa, Fergusson:2014tza, Ade:2015lrj, Akrami:2018odb}), not even in combined analyses of the power spectrum and bispectrum~\cite{Fergusson:2014hya, Fergusson:2014tza, Meerburg:2015owa, Akrami:2018odb}, restricting the feature amplitudes to the percent level relative to the scalar amplitude~$A_\mathrm{s}$. The application of the previously mentioned reconstruction techniques to CMB~data also points to a featureless power spectrum over the accessible range of scales and within current error bars~\cite{Dvorkin:2010dn, Dvorkin:2011ui, Planck:2013jfk, Achucarro:2014msa, Ade:2015lrj, Aghamousa:2017uqe, Obied:2017tpd, Akrami:2018odb, Obied:2018qdr}. Having said that, there are a couple of interesting candidates of marginal statistical significance~\cite{Ade:2015lrj, Akrami:2018odb}. These include the dip in power in the temperature power spectrum around multipoles of $\ell \sim 20 - 40$ and another oscillatory feature around $\ell \sim 700 - 800$. Future polarization data will be able to reduce the error bar by a factor of two for the mentioned low-$\ell$ feature candidate~\cite{Miranda:2014fwa}.

\vspace{3pt}
\textbf{Optical Galaxy Surveys} are the current frontier in the search for oscillations and are expected to improve the constraints significantly (cf.~\cite{Chantavat:2010vt, Chen:2016vvw, Ballardini:2016hpi, LHuillier:2017lgm, Palma:2017wxu, Ballardini:2017qwq, Beutler:2019prep} for forecasts). Spectroscopic galaxy surveys can probe very large volumes and have a full three-dimensional sampling of the underlying density field, which means that the maximum oscillation frequency is limited entirely by the volume of the survey -- the bigger the survey, the smaller the fundamental frequency and, consequently, the higher the maximal~$\omega_{\hskip1pt\mathrm{lin}}$ that can be constrained. The biggest drawback is that the usable range of scales is limited to those that remain in the linear and weakly non-linear regime. Having said that, non-linear corrections still have to be correctly accounted for~\cite{Vlah:2015zda, Beutler:2019prep}. We however do not need to model the full shape of the power spectrum, but only the oscillatory part, which makes it a somewhat easier problem than the full non-linear treatment of biased tracers.

The current best limits inferred from galaxy clustering data of the Baryon Oscillation Spectroscopic Survey~(BOSS) alone are competitive with those derived from current Planck CMB~data for the accessible range of feature frequencies~\cite{Beutler:2019prep} and will improve by orders of magnitude with future surveys. (It is of course natural to combine CMB~and LSS~data in the feature search which has in particular been explored in~\cite{Hu:2014hra, Benetti:2016tvm, Zeng:2018ufm, Beutler:2019prep}.) In photometric surveys, the large radial kernels for weak lensing and galaxies with photometric errors smear the signal on most scales. On the largest scales probed by these surveys, they can however remain competitive due to the raw number of objects which can be several orders of magnitude larger than what can currently be achieved in spectroscopic surveys. We also note that LSST- and Euclid-like experiments will be able to reduce the error bar by a factor of five for the mentioned high-$\ell$ candidate in the CMB~\cite{Chen:2016vvw, Ballardini:2016hpi}.

\vspace{3pt}
\textbf{Future 21\,cm Surveys}, which operate at high redshifts, such as the recently proposed Stage~\textsc{ii} experiment~\cite{Ansari:2018ury}, hold the promise to improve the constraints by another few orders of magnitude~\cite{Chen:2016zuu, Xu:2016kwz}. In particular, there is three times more comoving volume available in the redshift range $z=2-6$ compared to $z<2$. More importantly, the universe is more linear and the tracer less biased, which allows an increase by a factor of about two in the maximum wavenumber used to search for these features.

\vspace{3pt}
\textbf{Spectral Distortions} of the CMB black body spectrum provide an entirely complementary window on the primordial power spectrum and small-scale features since they are uniquely sensitive to the primordial amplitude at scales of $k \simeq \SIrange{1}{e4}{\per\Mpc}$ (cf.~e.g.~\cite{Barnaby:2009dd, Chluba:2015bqa}). An experiment like PIXIE~\cite{Kogut:2011xw} or PRISM~\cite{Andre:2013nfa} could set interesting constraints on departures from a featureless primordial power spectrum in this range which is inaccessible in the~CMB and challenging to reliably observe in~LSS.

\vspace{3pt}
As discussed above, the sensitivity to a general feature model is difficult to forecast since different models lead to different fiducial templates. One possibility to simultaneously visualize constraints from various probes and models is to imagine a \emph{feature spectrometer}, i.e.~considering the sensitivity to a linear feature model and decomposing any other feature into a sum of linear oscillations. We note that this picture has limitations, particularly for features localized in $k$-space, since the feature templates are not random fields, but instead have well-defined shapes (or phase relations in decomposition). With this caveat in mind, we show forecasts for linear features in the right panel of Fig.~\ref{fig:features} which demonstrates a beautiful synergy between CMB and LSS~experiments. LSS~observations have a smaller dynamical range in the feature frequency~$\omega_{\hskip1pt\mathrm{lin}}$ for two reasons: the largest available scales in real space are intrinsically smaller since a comoving scale per radian is considerably larger at the surface of last scattering, and the range of scales available from the fundamental mode to the onset of non-linear evolution is also smaller. On the other hand, over the range of scales in which both observational probes are sensitive, LSS~surveys are appreciably more sensitive which is a direct result of a three-dimensional, rather than two-dimensional sampling of the density fluctuations.

\section{Conclusions}

The main take-home points of this white paper are as follows:

\begin{itemize}[leftmargin=0.8cm]
\setlength\itemsep{-.05cm}
\item In theoretical attempts to connect the inflationary modeling to fundamental physics, departures from the minimal power-law power spectrum of initial fluctuations are ubiquitous.

\item Given the lack of our understanding of fundamental physics, there are no useful priors on the scale or amplitude of these features. We should therefore consider as much of parameter space that is amenable for cosmological searches.

\item The~CMB will dominate the sensitivity for the largest feature frequencies, while LSS surveys will keep improving the sensitivity elsewhere. The total survey volume, which determines the largest available scale, and the total number of linear and quasi-linear modes that preserve the primordial information are very good proxies for the survey sensitivity of such searches.
\end{itemize}

\clearpage
\makeatletter
\renewcommand\section{\@startsection{section}{1}{\z@}%
	{-3.5ex \@plus -1.3ex \@minus -.7ex}%
	{2.3ex \@plus.4ex \@minus .4ex}%
	{\normalfont\large\bfseries}}
\makeatother
\bibliographystyle{utphys}
\bibliography{whitepaper}
\end{document}